\documentclass[a4paper,11pt]{article}
\usepackage{pos}

\title{High-resolution Imaging of Two Radio Quasars at the End of the Reionization Epoch}
\ShortTitle{Two Quasars at the End of Reionization}

\author*[a]{Krisztina~Perger}
\author[b]{Yingkang~Zhang}
\author[a,c]{S\'andor~Frey}
\author[b]{Tao~An}
\author[d,e,a]{Krisztina~\'E.~Gab\'anyi}
\author[f,g]{Leonid~I.~Gurvits}
\author[h]{Chorng-Yuan~Hwang}
\author[h]{Ekaterina~Koptelova}
\author[f]{Zsolt~Paragi}
\author[b]{Ailing~Wang}

\affiliation[a]{Konkoly Observatory, ELKH Research Centre for Astronomy and Earth Sciences (MTA Centre of Excellence), Konkoly Thege Mikl\'os \'ut 15-17, H-1121, Budapest, Hungary}
\affiliation[b]{Shanghai Astronomical Observatory,  CAS, 80 Nandan Road, Shanghai 200030, China}
\affiliation[c]{Institute of Physics and Astronomy, ELTE E\"otv\"os Lor\'and University, P\'azm\'any P\'eter s\'et\'any 1/A, H-1117, Budapest, Hungary}
\affiliation[d]{Department of Astronomy, Institute of Physics and Astronomy, ELTE E\"otv\"os Lor\'and University,  P\'azm\'any P\'eter s\'et\'any 1/A, H-1117, Budapest, Hungary}
\affiliation[e]{ELKH-ELTE Extragalactic Astrophysics Research Group, E\"otv\"os Lor\'and University,  P\'azm\'any P\'eter s\'et\'any 1/A, H-1117, Budapest, Hungary}
\affiliation[f]{Joint Institute for VLBI ERIC, Oude Hoogeveensedijk 4, 7991 PD Dwingeloo, The Netherlands}
\affiliation[g]{Faculty of Aerospace Engineering, Delft University of Technology, Kluyverweg 1, 2629 HS Delft, The~Netherlands}
\affiliation[h]{Graduate Institute of Astronomy, National Central University,  Taoyuan City, 32001, Taiwan}

\emailAdd{perger.krisztina@csfk.org}
\emailAdd{ykzhang@shao.ac.cn}
\emailAdd{frey.sandor@csfk.org}
\emailAdd{antao@shao.ac.cn}
\emailAdd{k.gabanyi@astro.elte.hu} 
\emailAdd{lgurvits@jive.eu}
\emailAdd{hwangcy@astro.ncu.edu.tw}
\emailAdd{koptelova@astro.ncu.edu.tw}
\emailAdd{zparagi@jive.eu}
\emailAdd{wal@shao.ac.cn}

\abstract{There are approximately 250 quasars discovered at redshift $z\geq6$, of which only a handful were detected in radio bands, and even fewer were imaged with the highest resolution very long baseline interferometry (VLBI) technique. Here we report the results of our dual-frequency observations with the Very Long Baseline Array (VLBA) of two such recently discovered quasars, VIKING J231818.35$-$311346.3 at $z=6.44$ and FIRST J233153.20$+$112952.11 at $z=6.57$. Both extremely distant sources were imaged with VLBI for the first time. The radio properties of the former are consistent with those of quasars with young radio jets. The latter has an UV/optical spectrum characteristic of BL Lac objects, of which no others have been found beyond redshift 4 so far. Our VLBA observations revealed a flat-spectrum compact radio source. }

\FullConference{%
  15th European VLBI Network Mini-Symposium and Users' Meeting (EVN2022)\\
  11-15 July 2022\\
  University College Cork, Ireland
}

\begin{document}
\maketitle

\section{Introduction}
To date, there are approximately 250 quasars identified at redshifts $z > 6$, of which only around a dozen were found to have a counterpart in radio wavebands, and only a handful were imaged with the highest resolution very long baseline interferometry (VLBI) technique  \cite{2017FrASS...4....9P}: \text{NDWFS~J14276$+$3312} \cite{2008AJ....136..344M,2008A&A...484L..39F}, CFHQS~J1429$+$5447 \cite{2011A&A...531L...5F}, SDSS~J010013.02$+$280225.92 \cite{2017ApJ...835L..20W}, PSO~J030947.49$+$271757.31 \cite{2020A&A...643L..12S}, and PSO~J172.3556$+$18.7734 \cite{2021AJ....161..207M}. Increasing the number of observations targeting such high-redshift objects is crucial for achieving better understanding of the formation of the earliest supermassive black holes and the cosmological evolution of the astrophysical objects associated with them.  Here we report results of the first VLBI imaging observations of two recently discovered extremely high redshift quasars. 

\section{VIKING J231818.35$-$311346.3}
The source VIKING J231818.35$-$311346.3 (J2318$-$3113) was identified as a quasar at redshift $z=6.44$ \cite{2018ApJ...854...97D}, and detected in the radio domain at 888~MHz in the Galaxy and Mass Assembly and Rapid ASKAP Continuum Surveys \cite{2021A&A...647L..11I}. J2318$-3113$ was found to be a radio-loud  (radio loudness parameter $R=70$, \cite{2021A&A...647L..11I}) steep spectrum quasar, with a spectral index of $\alpha=-1.24$  between 888~MHz and 5.5~GHz (\cite{2022A&A...663A..73I}, following the $S \sim\nu^\alpha$ convention, where $S$ is the flux density and $\nu$ is the frequency).

Our 4.7 and 1.6~GHz Very Long Baseline Array (VLBA) observations were conducted on August 2 and 16, 2021, respectively (project code: BZ083, P.I. Y. Zhang). The raw data were correlated at the Distributed FX (DiFX) correlator \cite{2011PASP..123..275D} in Socorro (New Mexico, USA), with an integration time of 2~s and a recording data rate of 2~Gbps. Each 5~h observation was carried out in phase-referencing mode, using J2314$-$3138 as a phase calibrator. The data were recorded in 512~MHz frequency bands. The correlated data sets were calibrated using the US National Radio Astronomy Observatory (NRAO) Astronomical Image Processing System (\textsc{aips}) software package \cite{2003ASSL..285..109G}. The amplitude and phase self-calibration, imaging, and model fitting were carried out with \textsc{difmap} program \cite{1994BAAS...26..987S}. A detailed description of the observations, calibrations, and model fitting is given in the original paper describing the project \cite{2022A&A...662L...2Z}. 

\begin{figure}[h!]
\centering
\includegraphics[width=0.45\linewidth]{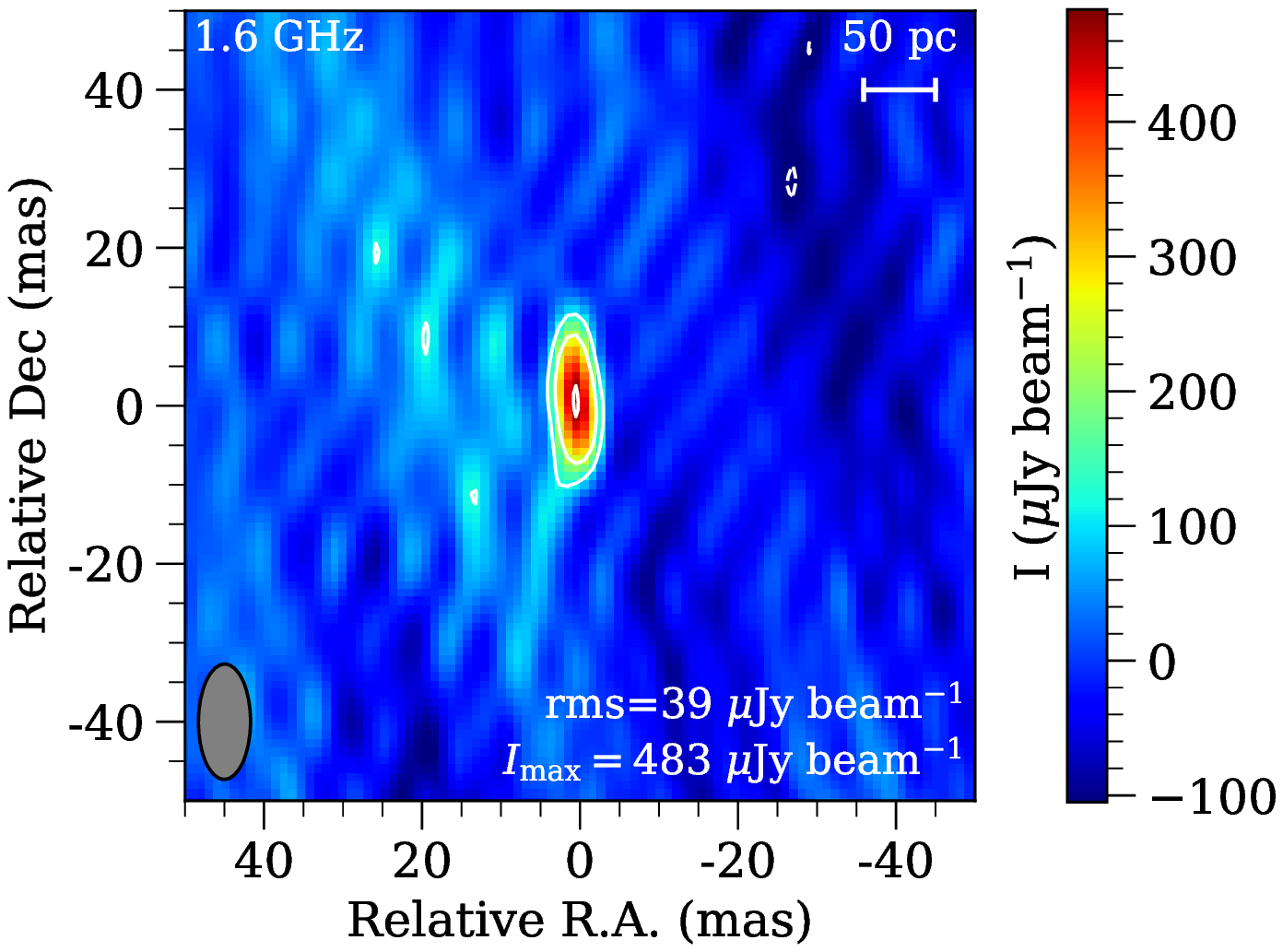}~~
\includegraphics[width=0.45\linewidth]{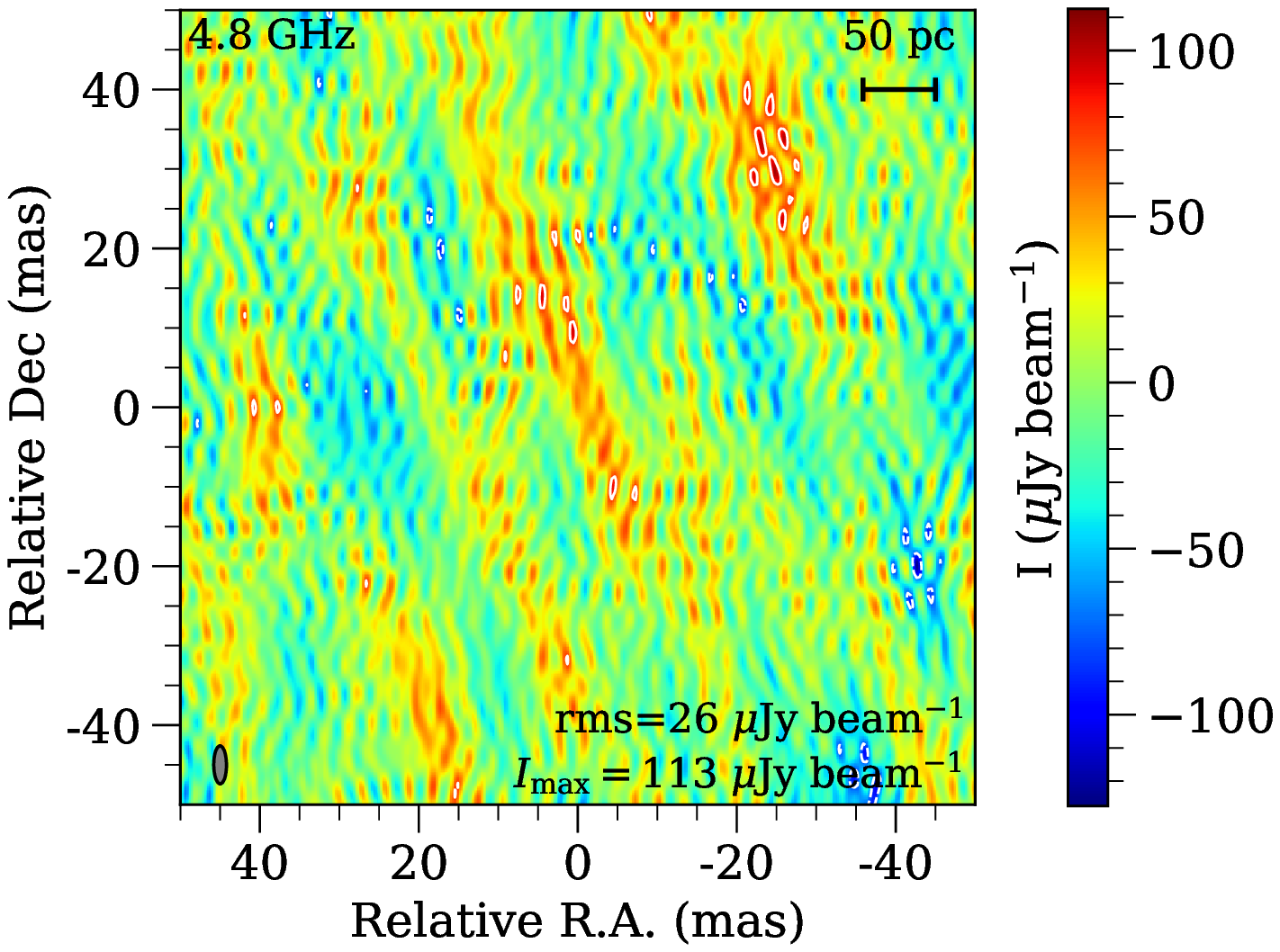}
\caption{Naturally weighted images obtained from the VLBA observations of J2318$-$3113. The images are reproduced from Fig. 2 of \cite{2022A&A...662L...2Z}. The intensity colour scales are on the right-hand sides of the panels. The lowest contours start at $\pm3\sigma$ rms noise, and the positive levels increase by a factor of 2. The restoring beams are shown in the lower left corners. The beam sizes of the 1.6~GHz  and 4.8~GHz maps are $14.6$~mas$\times4.6$~mas and $4.8$~mas$\times1.6$~mas, with position angles of $0.3^\circ$ and $-0.3^\circ$ (measured from North through East), respectively.}\label{fig:vik}
\end{figure}

The resulting radio images are shown in Fig.~\ref{fig:vik}. The source was detected as a compact `core' at 1.6~GHz with a flux density of 550~$\mu$Jy, while no compact component brighter than $130~\mu$Jy (i.e. $5\sigma$ image noise) was detected at 4.7~GHz \cite{2022A&A...662L...2Z}. This constrains the spectral index of the pc-scale feature to $\alpha<-1.2$, which is consistent with the value found for the kpc-scale radio emission \cite{2022A&A...663A..73I}. 
Detailed analysis \cite{2022A&A...663A..73I} showed that the radio emission of J2318$-$3113 likely originates from a recently formed quasar jet, as fitting both a curved (peaked) and a double power law model to the radio spectrum implies a young kinetic ($\sim500$ yr) and radiative age ($\lesssim9\times10^4$ yr).

\section{FIRST J233153.20$+$112952.11}
The observed near-infrared spectrum of the quasar FIRST J233153.20$+$112952.11 (J2331$+$1129) shows no detected emission lines \cite{2022ApJ...929L...7K}. Its redshift was estimated based on the  Gunn–Peterson trough found at $0.921~\mu$m, giving a lower limit on the redshift of $z=6.57$. The optical-to-radio spectral energy distribution of J2331$+$1129 is dominated by the synchrotron emission of the jet and closely resembles those of BL Lac objects. J2331$+$1129 has a flat radio spectrum ($\alpha=-0.01$) and non-thermal UV/optical continuum of spectral index $\alpha_{\rm opt}=1.43\pm0.23$. The UV/optical and radio fluxes of J2331$+$1129 vary on timescales of months to years. The spectral properties indicate that this quasar might be the highest-redshift BL Lac object discovered to date, of which none has been found so far beyond $z>4$ \cite{2022ApJ...929L...7K}.

We conducted observations of J2331$+$1129 with the VLBA at 1.6 and 4.8~GHz on February  1 and 4, 2022 (project code: BF132, P.I.: S. Frey). The phase-referencing observations were conducted for 4 hours at each frequency, with 2~Gbps and 4~Gbps data rates for the 1.6 and 4.8 GHz observations, respectively, using J233040.85$+$110018.7 as a phase calibrator. The raw data recorded at each telescope were correlated
at the DiFX correlator \cite{2011PASP..123..275D} in Socorro (New Mexico, USA). For the 1.6 and 4.8~GHz data, the observations were conducted in 2 and 4 intermediate frequency channels (IFs),  respectively. The total bandwidth was 128 MHz for both frequency bands. Phase and amplitude calibration, self-calibration, imaging, and model fitting were carried out in the same manner as for  J2318$-$3113, and will be described in detail in \cite{freyinprep}.

\begin{figure}
\centering
\includegraphics[width=0.45\linewidth]{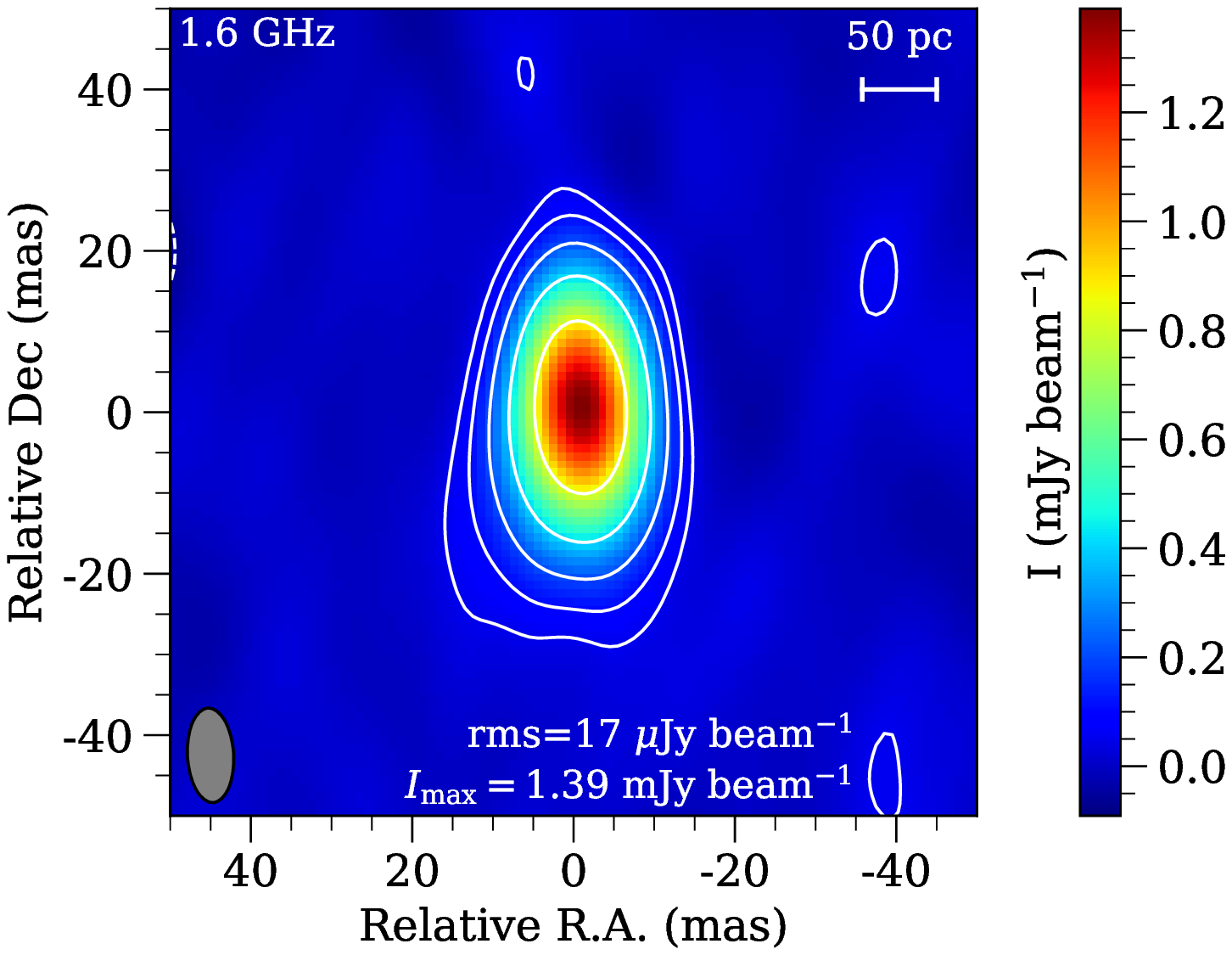}~~
\includegraphics[width=0.45\linewidth]{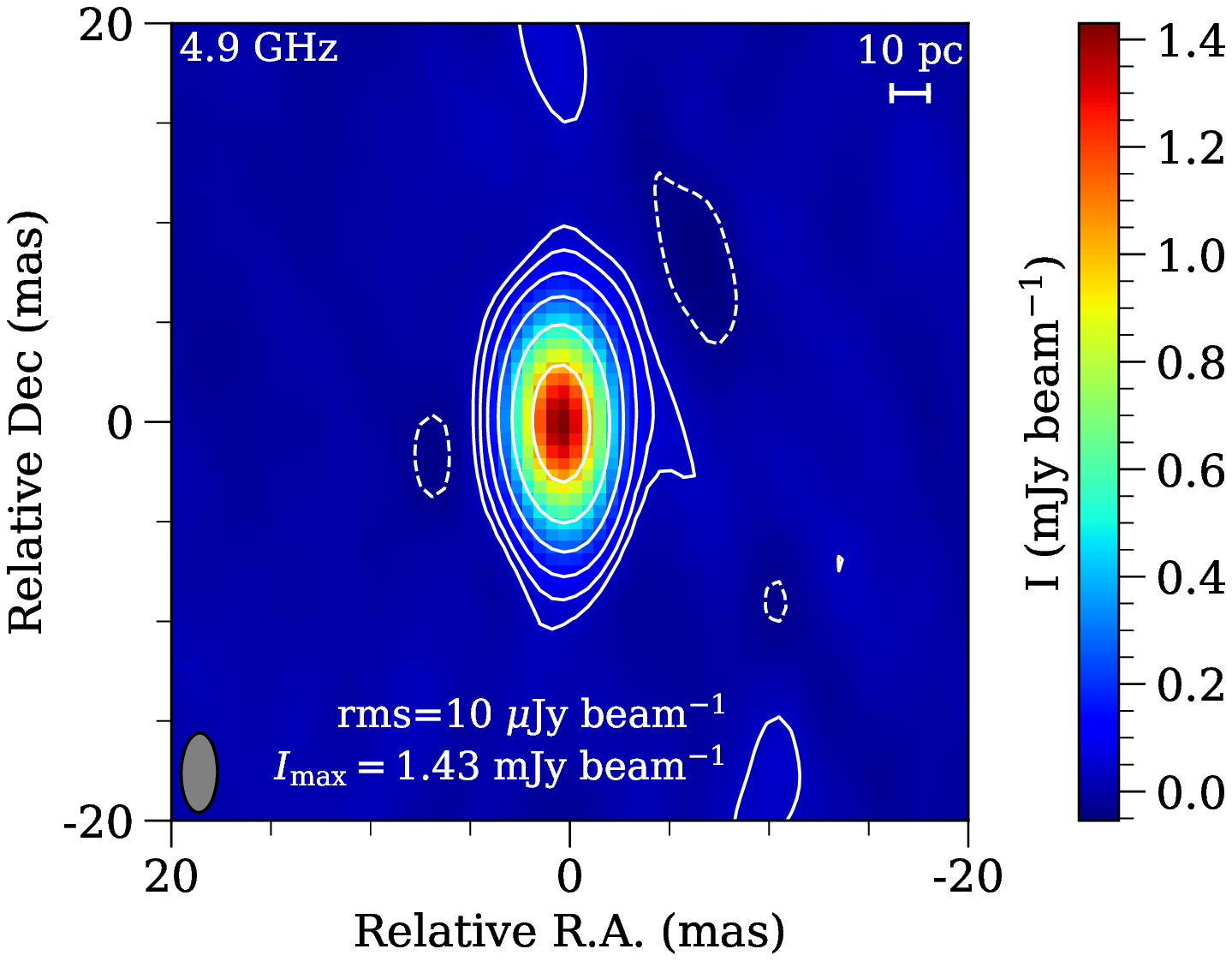}
\caption{Naturally weighted VLBA images of the source  J2331$+$1129. The intensity colour scales are on the right-hand sides of the panels. The lowest contours start at $\pm3\sigma$ rms noise, and the positive levels increase by a factor of 2. The restoring beams are shown in the lower left corners.  The beam sizes of the 1.6~GHz  and 4.9~GHz maps are $11.7$~mas$\times5.7$~mas and $4.0$~mas$\times1.8$~mas, with position angles of $3.2^\circ$ and $-1.0^\circ$ (measured from North through East), respectively.}\label{fig:first}
\end{figure}

The quasar J2331$+$1129 was detected at both frequencies. We found that the source is compact (Fig.~\ref{fig:first}), with flux densities $S_\mathrm{1.6GHz}=1.8~$mJy and $S_\mathrm{4.8GHz}=1.6$~mJy, indicating a flat spectrum ($\alpha_\mathrm{pc}=-0.11$). The brightness temperature values ($T_\mathrm{b}\sim10^8-10^9$~K) confirm the non-thermal nature of the radio emission. To facilitate a comparison with other sources, we calculated the 1.4~GHz rest-frame radio power as $P=4\pi S D_\mathrm{L}^2(1+z)^{-\alpha-1}$, where $D_\mathrm{L}$ is the luminosity distance of the source. The high value of $P=1.4\times 10^{26}~$W~Hz$^{-1}$ suggests that  the emission originates from a powerful active galactic nucleus (see \cite{2019MNRAS.490.2542P} and references therein), provided that the source is indeed at such a high redshift.

We found no evidence for strongly Doppler-boosted radiation, as the brightness temperature does not reach the equipartition limit of $T_\mathrm{b,eq} \approx 5 \times 10^{10}$~K \cite{1994ApJ...426...51R}, usually assumed as the intrinsic value. Such an enhancement with Doppler factor $\delta = T_\mathrm{b}/T_\mathrm{b,eq}> 1$ would be expected from the jet of a flat-spectrum radio quasar or a low-synchrotron-peaked BL Lac object. The less strongly beamed jets with Doppler factors $\delta = T_\mathrm{b}/T_\mathrm{b,eq}\lesssim1$ are more typical for intermediate- and high-synchrotron-peaked BL Lac objects \cite{2012ApJ...757...25L}. The relatively low measured brightness temperature values could also imply that the particles and the magnetic field are not in equipartition in the source.

\section*{Acknowledgements}
This research was funded by the National SKA Program of China (grant No. 2022SKA0130103) and the China–Hungary project funded by the Ministry of Science and Technology of China and Chinese Academy of Sciences. YKZ was sponsored by Shanghai Sailing Program under grant number 22YF1456100. This research was supported by the Hungarian National Research, Development and Innovation Office (NKFIH), grant number OTKA K134213. The National Radio Astronomy Observatory is a facility of the National Science Foundation operated under cooperative agreement by Associated Universities, Inc.


\begin{thebibliography}{99}
\bibitem[\protect\citeauthoryear{Decarli et al.}{2018}]{2018ApJ...854...97D} Decarli R., Walter F., Venemans B.~P., Ba{\~n}ados E., Bertoldi F., Carilli C., Fan X., et al., 2018, ApJ, 854, 97. doi:10.3847/1538-4357/aaa5aa

\bibitem[\protect\citeauthoryear{Deller et al.}{2011}]{2011PASP..123..275D} Deller A.~T., Brisken W.~F., Phillips C.~J., Morgan J., Alef W., Cappallo R., Middelberg E., et al., 2011, PASP, 123, 275. doi:10.1086/658907

\bibitem[\protect\citeauthoryear{Frey et al.}{2008}]{2008A&A...484L..39F} Frey S., Gurvits L.~I., Paragi Z.,  Gab{\'a}nyi K.~{\'E}., 2008, A\&A, 484, L39. doi:10.1051/0004-6361:200810040

\bibitem[\protect\citeauthoryear{Frey et al.}{2011}]{2011A&A...531L...5F} Frey S., Paragi Z., Gurvits L.~I., Gab{\'a}nyi K. {\'E}., Cseh D., 2011, A\&A, 531, L5. doi:10.1051/0004-6361/201117341

\bibitem[\protect\citeauthoryear{Frey et al.}{(in preparation)}]{freyinprep} Frey S. et al. 2023, A\&A, in preparation

\bibitem[\protect\citeauthoryear{Greisen}{2003}]{2003ASSL..285..109G} Greisen E.~W., 2003, in Information Handling in Astronomy -- Historical Vistas, ed. A. Heck, Astrophysics and Space Science Library (Dordrecht: Kluwer), 285, 109. doi:10.1007/0-306-48080-8\_7

\bibitem[\protect\citeauthoryear{Ighina et al.}{2021}]{2021A&A...647L..11I} Ighina L., Belladitta S., Caccianiga A., Broderick J.~W., Drouart G., Moretti A., Seymour N., 2021, A\&A, 647, L11. doi:10.1051/0004-6361/202140362
\bibitem[\protect\citeauthoryear{Ighina et al.}{2022}]{2022A&A...663A..73I} Ighina L., Leung J.~K., Broderick J.~W., Drouart G., Seymour N., Belladitta S., Caccianiga A., et al., 2022, A\&A, 663, A73. doi:10.1051/0004-6361/202142733

\bibitem[\protect\citeauthoryear{Koptelova \& Hwang}{2022}]{2022ApJ...929L...7K} Koptelova E., Hwang C.-Y., 2022, ApJL, 929, L7. doi:10.3847/2041-8213/ac61e0

\bibitem[\protect\citeauthoryear{Linford et al.}{2012}]{2012ApJ...757...25L} Linford J.~D., Taylor G.~B., Schinzel F.~K., 2012, ApJ, 757, 25. doi:10.1088/0004-637X/757/1/25

\bibitem[\protect\citeauthoryear{Momjian, Carilli, \& McGreer}{2008}]{2008AJ....136..344M} Momjian E., Carilli C.~L., McGreer I.~D., 2008, AJ, 136, 344. doi:10.1088/0004-6256/136/1/344

\bibitem[\protect\citeauthoryear{Momjian et al.}{2021}]{2021AJ....161..207M} Momjian E., Ba{\~n}ados E., Carilli C.~L., Walter F., Mazzucchelli C., 2021, AJ, 161, 207. doi:10.3847/1538-3881/abe6ae

\bibitem[\protect\citeauthoryear{Perger et al.}{2017}]{2017FrASS...4....9P} Perger K., Frey S., Gab{\'a}nyi K. {\'E}., T{\'o}th L.~V., 2017, FrASS, 4, 9. doi:10.3389/fspas.2017.00009

\bibitem[\protect\citeauthoryear{Perger et al.}{2019}]{2019MNRAS.490.2542P} Perger K., Frey S., Gab{\'a}nyi K. {\'E}., T{\'o}th L.~V., 2019, MNRAS, 490, 2542. doi:10.1093/mnras/stz2723

\bibitem[\protect\citeauthoryear{Readhead}{1994}]{1994ApJ...426...51R} Readhead A.~C.~S., 1994, ApJ, 426, 51. doi:10.1086/174038

\bibitem[\protect\citeauthoryear{Shepherd, Pearson, \& Taylor}{1994}]{1994BAAS...26..987S} Shepherd M.~C., Pearson T.~J., Taylor G.~B., 1994, BAAS, 26, 987

\bibitem[\protect\citeauthoryear{Spingola et al.}{2020}]{2020A&A...643L..12S} Spingola C., Dallacasa D., Belladitta S., Caccianiga A., Giroletti M., Moretti A., Orienti M., 2020, A\&A, 643, L12. doi:10.1051/0004-6361/202039458

\bibitem[\protect\citeauthoryear{Wang et al.}{2017}]{2017ApJ...835L..20W} Wang R., Momjian E., Carilli C.~L., Wu X.-B., Fan X., Walter F., Strauss M.~A., et al., 2017, ApJL, 835, L20. doi:10.3847/2041-8213/835/2/L20



\bibitem[\protect\citeauthoryear{Zhang et al.}{2022}]{2022A&A...662L...2Z} Zhang Y., An T., Wang A., Frey S., Gurvits L.~I., Gab{\'a}nyi K. {\'E}., Perger K., et al., 2022, A\&A, 662, L2. doi:10.1051/0004-6361/202243785

\end{thebibliography}
\end{document}